# A numerical study of self-focusing and guiding of laser pulse of duration shorter than plasma wavelength

D. Hazra[1*†] and A. Moorti[2,3]

[1]VIT Bhopal University, Bhopal-Indore highway, Kothrikalan, Sehore, Madhya Pradesh-466114, India

[2]Homi Bhabha National Institute, Training School Complex, Anushakti Nagar, Mumbai-400094, India

[3]Laser Plasma Division, Raja Ramanna Centre for Advanced Technology, Indore-452013, India

[*]Email:dhazraphys@gmail.com

[†]Corresponding Author: dhazraphys@gmail.com;

ORCHID ID: https://orcid.org/0000-0001-8726-4620

**Abstract:**

Self-focusing and guiding of ultra-short (pulse duration: L<$\lambda_P$: plasma wavelength), intense laser pulses in underdense plasma relevant to laser wakefield electron acceleration has been studied numerically. The analysis considers contribution of wakefield non-linearities along with relativistic effects. Stable propagation of laser pulse with mild periodic oscillations upto few Rayleigh lengths were observed. Parametric optimization of the different laser and plasma parameters performed show that, laser pulse as short as L/$\lambda_P$~0.42 is most stably guided upto few Rayleigh lengths for normalized laser intensity, $a_0$~0.9. Next effect of upward density ramps were also studied and it was found that inclusion of an upward density ramp not only enhances self-focusing and guiding of the ultra-short laser pulse but also reduces required minimum laser intensity compared to uniform density profile.

## 1. INTRODUCTION

Interaction and propagation of intense laser pulses in underdense plasma has been a subject of considerable interest [1, 2] due to various possible applications e.g. laser plasma electron acceleration [3]. Such studies usually require very high laser intensities ($\geq 10^{18}$ W/cm$^2$) at the interaction region, and are achieved by focusing laser pulses to a focal spot size of the order of few microns to few tens of microns using suitable focusing optics. For various applications like electron acceleration it is desirable to have long laser propagation distances inside underdense plasma. However, focused laser pulse propagation distance is limited to Rayleigh length given by, $Z_R = \pi r_0^2/\lambda$, (where $r_0$ is the initial laser spot radius at $1/e^2$ value of peak intensity and $\lambda$ is the laser wavelength) due to its natural diffraction. This relation is valid for ideal Gaussian beam, but in practical cases, the presence of higher order modes in the laser beam limits the beam from being perfect Gaussian. How much a real beam is deviated from the ideal Gaussian beam is described by $M^2$ parameter, which is defined as $W\theta = M^2 2\pi/\lambda$, where W and $\theta$ are the laser focal spot size and angular spread of the real beam. The $M^2$ parameter is a positive number ranging from unity to greater than unity: larger the value of $M^2$ larger is the deviation of real laser pulse from Gaussian. Considering $M^2$ parameter, Rayleigh length of a real beam is redefined as $Z_R' = \pi W^2/M^2\lambda$. Hence for larger values of $M^2$ Rayleigh length decreases and effectively the laser pulse is diffracted much earlier. For focal spot size of few microns to few tens of microns, the Rayleigh length ranges from sub mm to few mm. In addition, inside plasma, propagation distance is further limited due to ionization induced defocusing effects [4, 5]. The degree of ionization of an atom by an intense laser field depends on the atomic number (Z) of the atom and on the laser pulse intensity. Considering a Gaussian laser pulse profile both in the transverse and longitudinal direction, propagating inside a high Z gas target, the outer shell electrons having lower binding energies are ionized at the foot or the wings whereas, the inner shell electrons with comparatively higher binding energies are ionized at the peak of the laser pulse. Thus, corresponding to laser intensity profile, a radial density profile is formed with a density maximum at the peak of the pulse (i.e. on axis), subsequently reducing the on-axis refractive index of the plasma given by $\eta = [1-(\omega_p/\omega)^2]^{1/2} = (1-n_e/n_c)^{1/2}$, where $\omega$ is the laser frequency, $\omega_p(rad/sec) = (n_e e^2/m\varepsilon_0)^{1/2}$ is the plasma frequency, e is the electron charge, m is the mass of electron, $\varepsilon_0$ is the permittivity of free space, $n_e$ is the plasma electron density, $n_c(cm^{-3}) = 1.1\times10^{21}/\lambda^2(\mu m)$ is the critical density. This leads to increase in the on-axis phase velocity, given by $v_p=c/\eta$ (where c is the speed of light), compared to the wings, and hence acts like a diverging lens effect and results into laser pulse defocusing. Ionization induced defocusing effects can be minimized for low Z gas targets due to complete ionization occurring at the foot or wings of the laser pulse and no further ionization at the pulse peak. Therefore, to avoid the ionization induced defocusing effect low Z gases such as $H_2$ and He are used for laser plasma electron acceleration experiments. However without any guiding the laser propagation shall still be limited by Rayleigh length and is generally not sufficient to achieve GeV class acceleration of electrons which requires self-guiding and propagation of laser pulses about 1 cm or beyond.

Guiding and propagation of intense laser pulses in underdense plasma beyond Rayleigh lengths can be achieved by modification of the refractive index of the plasma such that the refractive index is maximum on the laser propagation axis compared to the wings i.e. $d\eta/dr<0$. Various non-linear processes such as relativistic self-focusing [6, 7], and ponderomotive channeling [11, 12] modify the refractive index of the plasma in such a way and hence helps self-guiding of the laser pulse over several Rayleigh lengths. At very high laser intensities, the electrons quiver relativistically in the laser field [6, 7]. The relativistic mass ($m'=\gamma m$, where $\gamma$ is the relativistic factor) effectively modifies the local plasma frequency ($\omega_p'=\omega_p/\sqrt{\gamma}$) and hence increases the refractive index along the laser axis, leading to relativistic self-focusing and guiding of the laser pulse [8-10]. Whereas, in case of ponderomotive channeling, the electrons are expelled radially outward from the laser propagation axis by the ponderomotive force of the laser pulse leading to increase in the refractive index along the laser axis [11, 12]. It has been shown that when laser power exceeds critical power for relativistic self-focusing, $P_c$ (GW)=17.4($n_c/n_e$), diffraction can be overcome, and relativistic self-guiding of the pulse occurs [6, 7]. Consideration of ponderomotive self-channeling along with relativistic self-focusing enhances this effect by slightly reducing the critical power for guiding to $P_c$ (GW) $\geq$ 16.2($n_c/n_e$) [11, 12].

Relativistic self-guiding of the laser pulse depends on laser and plasma parameters viz. laser pulse length (L) (L=$c\tau_L$, $\tau_L$ is laser pulse duration at FWHM: full width at half maximum), plasma wavelength ($\lambda_p$), initial laser spot radius ($r_0$), peak amplitude of the normalized laser intensity ($a_0$), input laser power (P) and plasma electron density ($n_e$). Theoretical reports on non-linear analysis of relativistic optical guiding have shown that long (L>$\lambda_p$) and broad ($r_0$>$\lambda_p$) laser pulses with slow rise times can be relativistically guided for P>$P_c$ [13-15]. Long pulse propagation is also affected by the wakefield generated at the foot of the laser pulse, which leads to self-modulation of the laser pulse and breaks the laser pulse envelop into beamlets of the order of $\lambda_p$ [1, 2, 16], which resonantly excites strong wakefields along with relativistic guiding of the laser pulse. Further in this case, it has also been reported that upward density ramp enhances self-focusing and guiding [17, 18-26], as due to increase of the plasma density, relativistic mass effect becomes much more pronounced [18].

However in the case of ultra-short laser pulses (L<$\lambda_p$) with faster rise times, earlier it was shown that due to the mutual cancellation of the plasma density response (transverse electron motion) and longitudinal motion the relativistic guiding effect reduces significantly [10-13]. Later, Lu *et al*. [27] through a phenomenological theory showed that in case of complete electron cavitation i.e. formation of bubble, ultra-short laser pulses ($L<\lambda_p$) having matched spot size for a given laser intensity and electron density can indeed be self-guided for more than one Rayleigh length, which was also verified through simulation and later in experiments also [28]. But in this case, laser propagation distance is limited by the pump depletion length [29]. It may be noted that, this regime of ultra-short laser pulse guiding is of particular interest for bubble regime of the laser wakefield electron acceleration (LWFA) [4-6, 28, 30-33]. However, a detailed analytical theory of ultra-short laser pulse guiding has not been provided. In this regard, Gorbunov *et al.* [34] studied effect of nonlinear processes such as relativistic effect and density perturbations due to generation of wakefields on the propagation of ultra-short ($L<\lambda_p$) and broad ($r_0>\lambda_p$) laser pulse. It showed that although the relativistic and longitudinal ponderomotive nonlinearities partly cancel each other the residual non compensated nonlinearities still has a significant effect to guide ultra-short laser pulses [34].

As discussed above, although detailed theoretical study on laser pulse propagation in the long pulse regime ($L>\lambda_p$) has been performed [1, 2, 11-16], there are not many reports on self-guiding of short laser pulses ($L<\lambda_p$) considering wakefield effects and also effect of density ramps has also not been considered. In this paper, we have numerically solved the problem of ultra-short laser pulse ($L<\lambda_P$) self-focusing and guiding in undersense plasma considering the formulation of Gorbunov *et al.* [34] taking into account the effect of wakefield. At first, with the motivation to find out a stable pulse propagation regime upto several Rayleigh lengths, we have performed a parametric study to understand the role of various laser and plasma parameters on ultra-short laser pulse guiding. It has been found that for pulses as short as $L/\lambda_p \sim 0.42$, a matched spot radius of 20 µm and a minimum laser intensity of $a_0 \sim 0.9$ are required for stable propagation. It is also shown that for guiding of $L<\lambda_p$ pulses, another additional criterion of $r_0 \geq \lambda_p$ must be satisfied. In addition, studies have also been further carried out with the inclusion of suitable optimized upward density ramp in the same formulation. Steepness of the density ramp was also varied and its effect on the stability of the pulse propagation was also taken into account in the study of optimization of the ramp. It was observed that for an optimized plasma density ramp, even for the short laser pulse ($L<\lambda_p$) case also upward density ramp enhances self-focusing and guiding. Not only that interestingly it was found that required minimum laser intensity is lowered to $a_0 \sim 0.5$ (in comparison to $a_0 \sim 0.9$ for uniform density profile) for stable propagation of ultra-short laser pulse associated with bubble formation. Hence inclusion of an upward plasma density ramp plays a crucial role in enhancing the self-focusing and also in lowering the laser intensity along the self-focusing distance, thereby compensating for pump depletion. Such a numerical optimization study of short ($L<\lambda_p$) Gaussian laser pulse propagation through underdense plasma considering wakefield effects with upward density ramp profile has not been reported yet and would help in maximizing the wakefield to design laser wakefield experiments.

In Section 2, we discuss the theoretical formulation of the study related to short laser ($L<\lambda_p$) propagation and guiding considering wakefield effects. In Section 3, we discuss the computational results of the formulation and finally in Section 4 we conclude our study in Conclusion.

## 2. A BRIEF OVERVIEW OF THEORY AND FORMULATIONS OF LASER PULSE FOCUSING AND GUIDING

### A. Long Pulse ($L>\lambda_p$) Without Wakefield:

First analytical treatment of relativistic self-focusing in underdense plasma was performed by Max et al. [6] evaluating thresholds for relativistic self-focusing given by $P_c$ (GW) = $17.4(n_c/n_e)$, which was also evaluated by others using analytical model [7, 35, 36]. Later, analytical solutions to z-dependent self-focusing problem were studied [7, 11, 12, 37-40]. In this case, two dimensional wave equation governing propagation of long laser pulses (L>$\lambda_p$) inside initial homogeneous plasma considering only relativistic self-focusing effect was derived. Laser spot size evolution was given by [7],

$$\frac{d^2R}{dz^2} = \frac{1}{Z_R^2 R^3}\left(1 - \frac{P}{P_c}\right) \quad (1)$$

Where, R=r/$r_0$ is the normalized spot radius, $r_0$ is the initial vacuum spot radius, $Z_R$ is the Rayleigh length and P/$P_c$=$k_p^2 a_0^2 r_0^2$/16 (for circular polarization). In Eq. (1) first term denotes vacuum diffraction and second term denotes relativistic self-focusing effect. By solving above equation it is shown that for P<$P_c$, the laser spot diffracts, for P=$P_c$ laser remains guided at r = $r_0$. Whereas, for P>$P_c$, further focusing occurs, and it can be seen that the laser spot size will collapse (R~0) at the self-focusing length of $Z_d$=$Z_R$/$\sqrt{(P/P_c-1)}$ [41, 42]. This is because considering $a_0^2$ << 1, $(1+a_0^2)^{-1/2}$ is approximated to only first order i.e. (1-$a_0^2$/2). However, consideration of higher order non-linearity prevents laser from collapsing [7, 12]. The smallest spot size to which the laser pulse is focused is given by, $\sigma = (a_0 n_c/n_e)^{1/2}\lambda/\pi$ [42].

Relativistic self-guiding has been also studied by deriving an envelope equation for the laser spot size [7, 12]. The generalised envelope equation permits analytically prediction of the evolution of the laser spot size, and also for a given P/$P_c$ a matched or equilibrium spot size can be easily obtained. The evolution of the radiation envelope is understood through the analogy of a single particle with orbit r(z) moving in a potential V(r) which is a function of normalised envelope radius x (=r/$a_0 r_0$) and P/$P_c$ on which the shape of the potential also depends and can be determined numerically. In general, for P>$P_c$, there exists a finite well with a minimum at x=$x_f$. The value of the laser focal spot corresponding to this $x_f$ gives the matching spot radius ($r_m$). With increase in P/$P_c$ the depth of the potential well increases and becomes narrower and the location of the minimum $x_f$ decreases (i.e. value of matching spot radius reduces). Hence it is possible to have a relativistic guiding of the laser with a stable laser spot size if initial laser spot size is same as the matching spot radius i.e. $r_0=r_m$ with initial plane wavefront. In case of $r_0<r_m$, the beam shall diffract and for $r_0>r_m$, the beam focuses. In the later case, with initial plane wavefront, the envelope will first converge to a minimum spot radius ($r_m$) and then oscillates between this minimum value and the initial spot radius value i.e. $r_0$ [7, 12]. However, in the case of highly converging wavefront the beam first focuses to a minimum spot and then expands indefinitely.

Subsequently, in addition to relativistic effect role of ponderomotive expulsion of the electrons by the laser field was also considered in various theoretical studies of propagation of long laser pulse ($L>\lambda_p$) in underdense plasma [11, 12, 15, 17, 37, 43-47]. Here, the consideration of the long ($L>\lambda_p$) pulse allows to assume a quasi-steady state and hence considering circular laser polarization the density perturbation equation is derived equating the radial electrostatic potential of the space charge field to the radial ponderomotive force. By solving paraxial wave equation, the normalized spot size of a Gaussian laser pulse was derived by Sprangle *et al*. [43] and is given by,

$$\frac{d^2R}{dz^2} = \frac{1}{z_R^2 R^3}\left(1 - \frac{P}{P_c} - \frac{a_0^2}{2R^2}\right) \qquad (2)$$

Here, the first term on the right hand side of Eq. (2) is the vacuum diffraction term, the second term is the relativistic self-focusing and the third term is the focusing due to ponderomotive transverse charge separation. Eq. (2) shows that, for $P/P_c<<1$ i.e., neglecting relativistic effects and considering $a_0<1$, the laser pulse will diffract as the magnitude of ponderomotive density modulation is not enough to focus the laser pulse. However, with $P>P_c$ and $a_0\sim1$, i.e. considering both relativistic and ponderomotive effects, the above equation shows that focusing of the pulse occurs earlier and with increased oscillation frequency compared to the case of only relativistic effect and the minimum laser spot size achieved is also reduced [12]. Further, consideration of both ponderomotive and relativistic self-focusing effects, reduces the threshold for self-guiding to $P_c$ (GW) $\geq 16.2(n_c/n_e)$ [11, 12] as compared to the only relativistic case [6].

Combined effect of relativistic and ponderomotive channelling was also studied using envelope equations in terms of single particle orbiting in a potential [12]. It was shown that potential curve in case of combined relativistic and ponderomotive effect is deeper, narrower, and matching laser spot size is reduced compared to the only relativistic case. It was shown that, for $P/P_c >1$ and given initial beam spot, the envelope i.e. (spot radius) will oscillate about an equilibrium beam radius with propagation distance. Also oscillation period was longer for only relativistic case compared to the combined relativistic and ponderomotive effect.

A different treatment of the problem of laser self-guiding under the combined influence of relativistic and ponderomotive effects has been provided by Tripathi *et al.* [46], where modification of dielectric constant ($\varepsilon$) of the plasma by the relativistic mass increase of the electrons quivering in high intense laser field and also due to the radial ponderomotive expulsion of the electrons was considered. A differential equation governing the evolution of the normalized spot radius (f=r/r$_0$) of the laser pulse with propagation distance (z) inside plasma is given by,

$$\frac{\partial^2 f}{\partial \zeta^2} = \frac{1}{f^3} - \frac{1}{2\varepsilon_0}\frac{\partial f}{\partial \zeta}\frac{\partial \varepsilon_0}{\partial \zeta} - \frac{Z_R^2}{r_0^2 \varepsilon_0}\phi f \qquad (3)$$

Here, $\zeta=Z/Z_R$ is the normalized propagation distance, and $\varepsilon_0$ and $\phi$ are the coefficients of $\varepsilon$ and in the paraxial approximation ($r^2 \ll r_0^2 f^2$), $\varepsilon$ is Taylor expanded and considered upto second order non-linearity. The first term of Eq. (3) is the vacuum diffraction term, second term is due to the plasma inhomogeneities and the third term is the non-linear self-focusing term considering relativistic and ponderomotive effects. Solving the above equation with boundary conditions that at z=0, f=1 and df/dz=0 i.e. assuming a plane wave initially, it was shown that with only the first term the pulse diffracts, whereas considering the first and the third term (relativistic and ponderomotive effect), the pulse gets focused and undergoes an oscillatory focusing and defocusing behavior along the propagation direction about the initial spot radius. Using same formulation effect of upward density ramp on laser self-focusing was also studied and it was shown that it enhances the effect of self-focusing [17, 47].

**B. Long Pulse (L>$\lambda_p$) With Wakefield:**

Wakefield effect was also considered and propagation of laser pulses in underdense plasma was formulated and numerically evaluated by Sprangle *et al.* [15, 16] in a weakly relativistic limit ($a^2$<<1) through self-consistent Maxwell fluid equations given by:

$$[\nabla_\perp^2 + \frac{2ik_0}{c}\frac{\partial}{\partial\tau}]a = k_p^2\left(N - \frac{|a|^2}{2}\right)a \qquad (4)$$

$$\frac{\partial^2 N}{\partial\xi^2} + k_p^2 N = \frac{\partial^2}{\partial\xi^2}\frac{|a|^2}{2} \qquad (5)$$

Here, $N = (n_e-n_0)/n_0$ is the normalized electron density perturbation; $k_0$ is the laser wave number. In the above case relativistic and wakefield effects are considered and transverse ponderomotive force effect was neglected assuming the broad laser pulse i.e. $r_0$>$\lambda_p$ and therefore a term $\nabla_\perp^2|a|^2/2$ was dropped from the right hand side of Eq. (5). In deriving the above equations independent variables $\xi$=ct-z and $\tau$=t were introduced, quasi-static approximations [13, 14] ($\partial/\partial\tau$=0) was applied in the fluid equation i.e. Eq. (4), and slowly varying envelop approximations was assumed i.e. |$\partial a/\partial\xi$|, |$\partial a/\partial c\tau$|<<|$k_0 a$|. The non-linear driving term (N-$a^2$/2) on the right hand side of Eq. (5) can be replaced by normalized electrostatic potential $\delta\phi$ such that

$$\left(N - \frac{a^2}{2}\right) = -\delta\phi \qquad (6)$$

And,

$$\left(\frac{\partial^2}{\partial\xi^2} + k_p^2\right)\delta\phi = k_p^2 a^2/2 \qquad (7)$$

And accordingly, the refractive index gets modified such that,

$$\eta = 1 - \left(\frac{\omega_p^2}{2\omega^2}\right)(1 - \delta\phi) \qquad (8)$$

For a long laser pulse with sufficiently smooth envelop, $\partial^2\phi/\partial\xi^2$ term in Eq. (7) can be neglected (which in turn neglects generation of plasma waves) and hence $\delta\phi=a^2/2$, which means refractive index modification is only by relativistic effects. Hence, long laser pulse ($L>\lambda_p$) can be relativistically self-guided. In addition inclusion of the effect of wakefield in the pulse propagation dynamics for long laser pulses ($L>>\lambda_p$) results in the self-modulation of laser pulse envelop [1, 2, 16].

### C. Short Pulse ($L<\lambda_p$) With Wakefield:

Using the above treatment Sprangle *et al*. [15] also showed that ultra-short laser pulse ($L<\lambda_p$) cannot be relativistically self-guided. As in case of ultra-short laser pulse ($L<\lambda_p$) with fast rise time, the pulse envelop becomes steep and compared to $\partial^2\phi/\partial\xi^2$ term, now $k_p^2$ term can be neglected in Eq. (7) such that for a constant intensity profile, the space charge potential $\phi=k_p^2 a_0^2 \xi^2/4$. This indicates that at the foot the laser pulse ($\xi=0$), $\phi=0$ and the refractive index given by Eq. (8) gets modified such that pulse suffers vacuum like diffraction. Thus, the cancellation of the plasma density response (transverse electron motion) and longitudinal motion significantly reduces the relativistic guiding effect in case of the ultra-short laser pulses [3]. Further, this fact was also verified through numerical simulation [15], where it was shown that ultra-short laser pulses ($L<\lambda_p$) are not guided although a small amount of guiding was seen at $L=\lambda_p$.

However, later Lu *et al*. [27] proposed a phenomenological theory of ultra-short ($L<\lambda_p$) laser pulse guiding in under dense plasma in conditions relevant to laser wakefield electron acceleration. It was shown that through complete electron cavitation or 'bubble formation' short laser pulses will be self-guided over Rayleigh length for a matched beam spot size determined by given laser intensity and plasma density. The matching condition for self-guiding of linear polarization laser pulse was given as:

$$k_P r_0 \approx k_P R \approx 2\sqrt{a_0}$$

$$\frac{r_0}{\lambda_P} = \frac{\sqrt{a_0}}{\pi} \qquad (9)$$

where R is the bubble radius. The results were further verified through simulation.

At the same time Gorbunov *et al*. [34] presented a study on the evolution of ultra-short ($L<\lambda_p$) and broad ($r_0>\lambda_p$) laser pulses in underdense plasma considering various non-linear effects in weakly relativistic ($a^2 \ll 1$) limit and described by a set of self-consistent equations:

$$2ik_0 \frac{\partial a}{\partial z} + \Delta_\perp a = k_P^2 \left(N - \frac{1}{4}|a|^2\right) a \qquad (10)$$

$$\frac{\partial^2 N}{\partial \xi^2} + k_P^2 N = \frac{1}{4}\left(\frac{\partial^2}{\partial \xi^2} + \Delta_\perp\right)|a|^2 \qquad (11)$$

In deriving Eq. (10) and (11), slowly varying envelop approximations was assumed and quasi-static approximations ($\partial/\partial\tau=0$) [13, 14] has been applied to the electron fluid equation. Here $\xi=ct-z$ measures the distance back from the head of the laser pulse moving with a group velocity ~c in the positive z direction. The first term on the right hand side of Eq. (10) describes the response of the plasma (wakefield) and the second term represents the contribution of first correction term to the relativistic factor. Here transverse nonlinearity due to ponderomotive force was also considered and therefore term $\nabla_\perp^2|a|^2/2$ also appears on the right hand side of Eq. (11) (c/f equation 5 above). It may be pointed out here that, in the earlier treatments by Sprangle *et al*., [16] transverse non-linearity was neglected for a broad ($r_0 > \lambda_p$) laser pulse, and it was shown that ultra-short laser pulses cannot be relativistically self-guided as the longitudinal non-linearity cancels the relativistic factor resulting in diffraction of the laser pulse. However, even in the case of broad pulse, consideration of finite pulse length effect which becomes significant in the case of ultra-short laser pulse compared to long laser pulse, the transverse non-linearity term in the right hand side of Eq. (11) cannot be neglected [34]. It was also shown that inclusion of the above finite transverse non-linearity (however small it may be) into the wave equation, effectively self-guides ultra-short laser pulses in under-dense plasma.

Assuming a Gaussian laser pulse final equation governing the spot size evolution of the laser pulse with the propagation distance is given by [34],

$$\frac{\partial^2 f}{\partial \zeta^2} = \frac{1}{f^3}\left\{1 - \alpha I(\xi)\left(1 + \frac{2}{\alpha f^2}\right)\right\} \tag{12}$$

where, the expression for $I(\xi)$ is given as,

$$I(\xi) = \frac{(a_0 \omega_P \tau_L)^2}{8\ln 2}\int_{-2\sqrt{\ln 2}\xi/c\tau_L}^{\infty} dx[1 - \text{erf}(x)] \tag{13}$$

Here, $\alpha = (k_p r_0)^2/8$ is the normalized initial squared spot size and $I(\xi)$ defines the probability integral. The first term on the right hand side of Eq. (12) is the vacuum diffraction term and the second term is the net focusing term generated due to various non-linearities involved. It was observed that some portion from the leading edge of the laser pulse will be diffracted and remaining part of the pulse would be self-focused and guided. The laser and plasma parameters decide a stability point ($\xi_c$) separating the diverging and converging parts in the pulse profile and a corresponding $I(\xi_c)$ given by [34],

$$I(\xi_c) = \frac{1}{(2+\alpha)} \qquad (14)$$

which decides how much part of the pulse will get focused and guided. Accordingly, a threshold laser intensity ($a_{0c}$) was given corresponding to the pulse centre ($\xi=0$) beyond which the self-focusing effect is applicable.

$$a_{0c} = \frac{1}{\omega_p \tau_L} \sqrt{\frac{4\ln 2}{1+(\frac{k_p r_0}{4})^2}} \qquad (15)$$

Eq. (12) is valid for only short laser pulses such that $k_p\xi<1$ i.e., upto which $\cos k_p(\xi-\xi')$ can be approximated to ~1. Also, the equation describes laser propagation and guiding in the weakly relativistic limits ($a_0<<1$), using linear wakefield terms. In deriving Eq. (12), laser energy loss i.e. pump depletion effect has not been considered. For typical laser (Ti: Sapphire laser at wavelength 800 nm and pulse duration of few tens of fs) and plasma parameters (few times $10^{18}$ cm$^{-3}$) relevant to wakefield excitation and electron acceleration experiments, pump depletion length (linear / non-linear) [27] is several times larger than applicable Rayleigh length and could be of the order of several mms [29]. Moreover in case of linear regime i.e. mildly relativistic regime, pump depletion effects would be minimal. Hence, a study of self-focusing and guiding behavior of the laser pulse inside underdense plasma without considering laser energy loss (pump depletion) could provide a reasonable understanding of the phenomena within the pump depletion length which could be several times $Z_R$. Several earlier theoretical studies report laser propagation and guiding [7, 11, 12, 15, 17, 38, 39, 47] and laser spot size evolution with the propagation distance to study the self-guiding behavior of the laser pulses without considering the pump depletion effects.

In the above formulation of short laser (L<$\lambda_p$) propagation and guiding we have further extended the study by including upward density ramp profiles into the equation. We have considered various upward density ramp profiles such as $n_e = n_0 \tan(z/d)$, where d is a dimensionless number which decides the steepness of the ramp profile and $n_0$ is the initial density, $n_e = n_0 + n_0 \tan(z/d)$, $n_e = n_0 + n_1 \tan(z/d)$ with $n_1 = 2n_0$ for our study and have also varied the steepness parameter d to study its effect. Density ramp profiles were included in the formulation through the plasma frequency $\omega_p = (n_e e^2/m\varepsilon_0)^{1/2}$. The modified plasma frequency and propagation vector on inclusion of the density ramp $n_e = n_0 \tan(z/d)$ is given by $\omega'_p = \left(\frac{e^2}{m\varepsilon_0}\right)^{1/2} \sqrt{n_0 \tan\left(\frac{z}{d}\right)}$ and $k'_p = \frac{2\pi c}{3.33 \times 10^{10}} \sqrt{n_0 \tan\left(\frac{z}{d}\right)}$. As a result accordingly Eq. (13), (14), (15) and finally equation governing the spot size evolution Eq. (12) gets modified. To study the laser propagation using upward density we have numerically solved the modified Eq. (12) with the modified electron density.

## 3. RESULTS AND DISCUSSIONS:

We have studied self-guiding and propagation of ultra-short ($L<\lambda_p$) laser pulse in underdense plasma for various laser and plasma parameters. In this regard, we have analyzed the above described analytical formulation of Gorbunov *et al*. [34] for ultra-short laser pulse propagation and have numerically solved the non-linear second order partial differential equation (i.e. Eq.12) governing the variation of normalized laser spot radius (f) with the propagation distance (z). Further, we have also included upward density ramp in solving Eq. (12) and studied the effect of various plasma density ramps on self-focusing and guiding of the laser pulse.

As discussed above, in case of ultra-short laser pulses ($L<\lambda_p$) guiding and propagation under influence of various non-linearities a part of the laser pulse is diffracted and remaining part is self-focused and guided [34]. How much fraction of the laser pulse will be diffracted and guided is determined by the laser and plasma parameters. To determine this, first we integrate Eq. (13) and plot $I(\xi)$ vs $\xi$ for a given laser parameters ($r_0$, $a_0$, and $\tau_L$) and plasma density ($n_e$). The function is a monotonically increasing function from leading front towards the trailing edge of the pulse. With increasing density the function becomes steeper towards the trailing end of the pulse. Now from Eq. (14) the value of $I(\xi_c)$ is obtained, and corresponding to that value of $\xi_c$ is obtained from the $I(\xi_c)$ vs $\xi$ graph which may lie between -1 (pulse trailing end) to +1 (pulse front end), and zero corresponds to the laser pulse center. The front part of the laser pulse upto the value of $\xi_c$ is diffracted and the trailing part of the pulse beyond $\xi_c$ is self-focused and guided. With the increase in laser intensity and/or plasma density, value of $\xi_c$ increases i.e. larger and larger part of the laser pulse is focused and guided. As discussed earlier, for a given laser and plasma parameter, an estimation of laser threshold intensity ($a_{0c}$) for self-guiding of ultra-short laser pulse beyond the pulse center can be estimated using Eq. (15), which decreases with increasing density. For example, Fig.1 shows the $I(\xi)$ vs $\xi$ plot for typical parameters of $n_e=3.5\times10^{18}$ cm$^{-3}$, $a_0=0.9$, $r_0=20$ μm and $\tau_L=25$ fs. For the above parameters, the value of $I(\xi_c)$ comes out to be 0.119 and the corresponding value of $\xi_c$ obtained is 0.48. This signifies that for the above mentioned parameters the part of the pulse for which $\xi>0.48$ will be diffracted and the remaining trailing part will get self-focused. For the above mentioned parameters, the threshold intensity $a_{0c}$ comes out to be ~0.309. The initial intensity $a_0=0.9$ used is much above this threshold intensity which leads to self-focusing and guiding of more than half of the fraction of laser pulse. Now, in order to study the propagation of laser pulse inside the plasma, Eq. (12) was solved numerically with boundary conditions that at z=0, f=1, df/dz=0 and evolution of the pulse centre ($\xi=0$) where the effect of non-linearity will be maximum was studied.

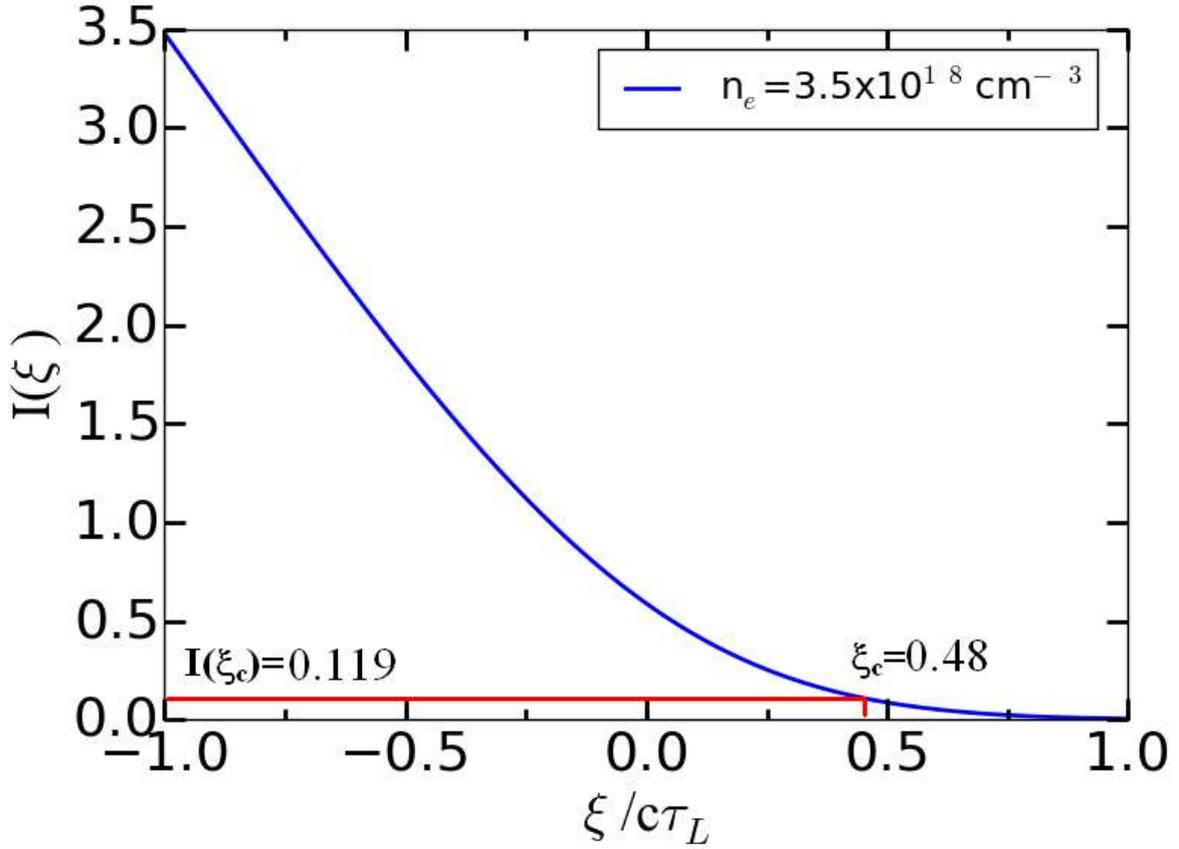

Fig.1. Variation of I(ξ) with $\xi/c\tau_L$

Fig. 2 shows the evolution of the normalized laser focal spot size 'f' with the normalized propagation distance '$Z/Z_R$' in the density range of $n_e=2.48\times10^{17}$ cm$^{-3}$ to $n_e=2\times10^{19}$ cm$^{-3}$ ($\lambda_p$=66.23-7.4 μm) for fixed $a_0$=0.9, $r_0$=20 μm and $\tau_L$=25 fs (L=7.5 μm). For this condition $P/P_c$ was in the range of ~0.18 to ~72.2. In fig. 2 it is seen that initial guiding of the laser pulse started at $n_e=3\times10^{18}$ cm$^{-3}$ ($P/P_c$=2.2) but occurrence of self-focusing of the laser pulse within $Z_R$ and stable propagation beyond several $Z_R$ with very few oscillations was observed at the density of $n_e=3.5\times10^{18}$ cm$^{-3}$ ($P/P_c$=2.56) i.e. $r_0/\lambda_p$~0.3 and $L/\lambda_p$~0.42. Here it may be pointed out that the above obtained value of density ($3.5\times10^{18}$ cm$^{-3}$) where stable guiding is observed is comparatively higher than the matching density $n_e=2.48\times10^{17}$ cm$^{-3}$ ($P/P_c$=0.18) as per the matching condition of Lu *et al*. following Eq. (9) [27]. Further, with increase in density tight focusing of the pulse with rapid oscillations takes place, which makes the propagation unstable. This is because of the fact that with increasing density the P/$P_c$ ratio increases leading to tight focusing of the pulse. As we increase the density to $n_e=2\times10^{19}$ cm$^{-3}$ (P/$P_c$=72.2) i.e. L/$\lambda_p$~1, the propagation is diverging and highly unstable signifying that the solution no longer holds in this regime as $\cos(k_p\xi)$ cannot assumed to be ~1.

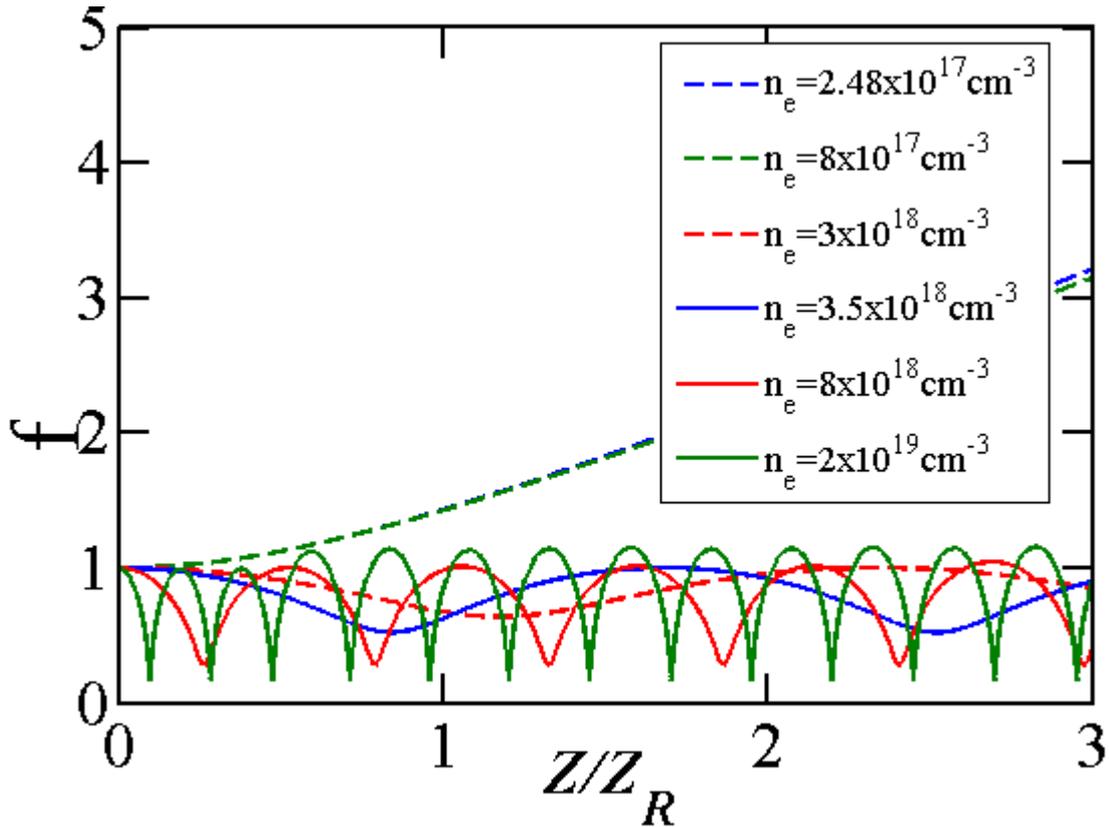

Fig.2. Evolution of f with normalized propagation distance (Z/$Z_R$) with varying electron density for fixed $a_0$=0.9, $r_0$=20 μm and $\tau_L$=25 fs.

Next, parametric study of stable guiding of the laser pulse was carried out for optimum plasma density of $n_e$ = 3.5×10$^{18}$ cm$^{-3}$ and L/$\lambda_p$ ~0.42. In Fig. 3 effect of varying laser intensity is shown for $a_0$=0.5 to 1 (P/$P_c$~0.8-3.15), at a constant density $n_e$=3.5×10$^{18}$ cm$^{-3}$, $r_0$=20 μm and $\tau_L$=25 fs. For lower laser intensities of $a_0$=0.5 (P/$P_c$=0.8), the pulse is diffracted. With slightly increasing $a_0$ to 0.7 (P/$P_c$=1.55), the pulse is guided with stable oscillations. At slightly higher laser intensity of $a_0=0.9$ ($P/P_c=2.56$), the pulse is focused within one $Z_R$ and stable guided beyond several $Z_R$ is observed. With further increase in $a_0=1$ ($P/P_c\sim3.15$), the pulse is more tightly focused with increase in the number of oscillations. So, the optimum laser intensity required for stable guiding is found to be in the range of $a_0 \sim 0.9$ ($P/P_c \sim 2.56$). Therefore, in addition to lowering the electron density, laser power P should be also increased simultaneously such that $P/P_c \geq 1$ and stable self-guiding of the ultra-short laser pulse takes place.

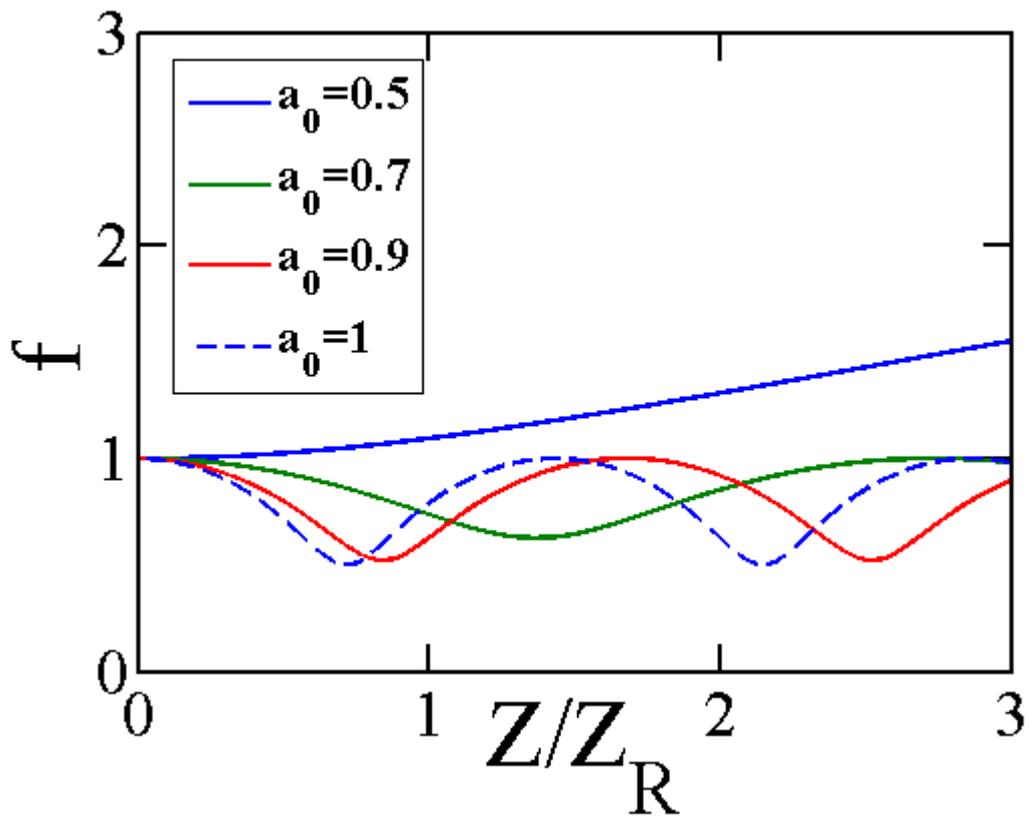

Fig.3. Evolution of f with normalized propagation distance ($Z/Z_R$) with varying laser intensity for fixed $n_e=3.5\times10^{18}$ cm$^{-3}$, $r_0=20$ μm and $\tau_L=25$ fs.

Next, in Fig. 4, effect of the initial laser spot radius on the propagation of ultra-short laser pulses is studied by varying spot radius from $r_0=5$ μm to $r_0=40$ μm for fixed parameters of $n_e=3.5\times10^{18}$ cm$^{-3}$ ($\lambda_p\sim17.64$ μm), $a_0=0.9$ and $\tau_L=25$ fs. For this condition $P/P_c$ was in the range of 0.16 to 10.21 and $r_0/\lambda_p$ ~0.28-2.26. We have observed that upto $r_0=10$ μm ($P/P_c=0.64$, $r_0/\lambda_p\sim0.56$) the pulse diffracts. For a spot radius of $r_0=20$ μm, ($P/P_c=2.56$, $r_0/\lambda_p\sim1.13$), self-focusing within one Rayleigh length and stable propagation beyond several $Z_R$ is observed. With further increase in value of $r_0$ to 30 - 40 μm ($P/P_c\sim5.73$-10.21, $r_0/\lambda_p\sim1.7$-2.26) the pulse gets more tightly focused with increasing number of oscillations. So, for ultra-short laser pulses, the condition $r_0\geq\lambda_p$ is favorable for pulse self-focusing and guiding whereas for $r_0<\lambda_p$ pulse diffracts. This is in agreement with the earlier report [28] where it was experimentally observed that guiding of ultra-short laser pulses requires long focal length focusing geometries providing $r_0>\lambda_p$ whereas for short focal length interactions ($r_0<\lambda_p$) beam break up prevents stable propagation of the pulse.

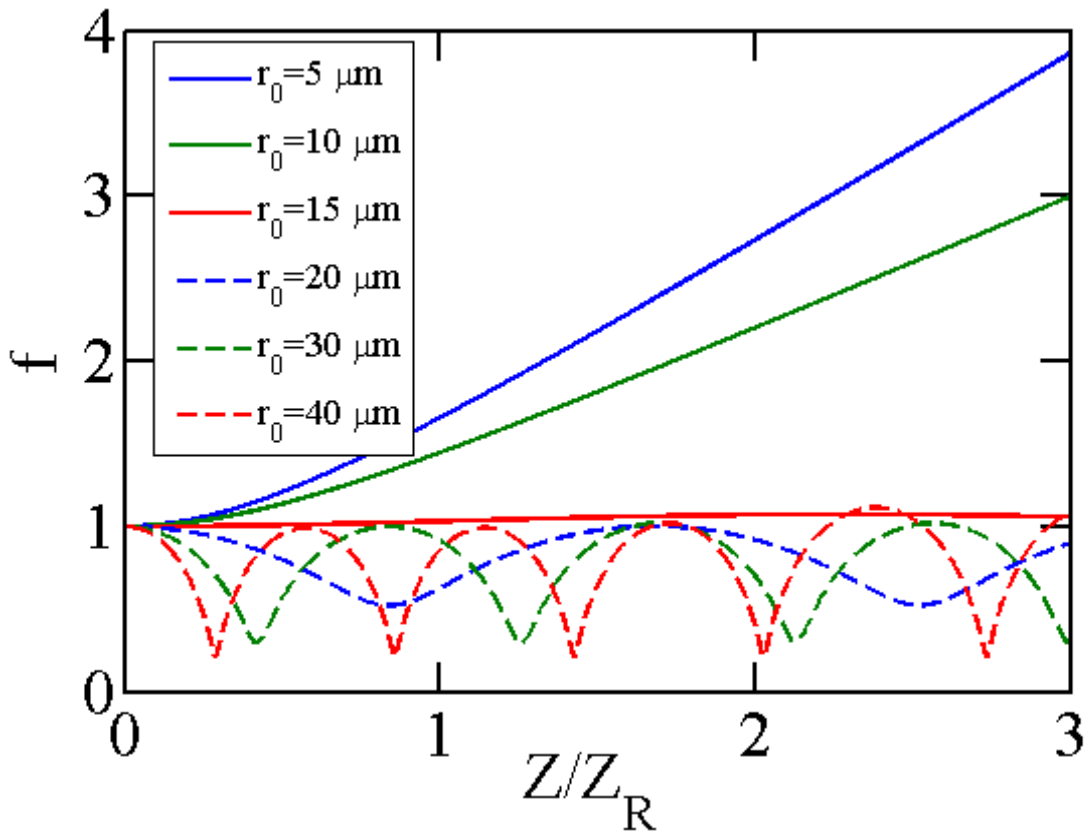

Fig.4. Evolution of f with normalized propagation distance ($Z/Z_R$) with varying laser spot size for fixed $n_e=3.5\times10^{18}$cm$^{-3}$, $a_0=0.9$ and $\tau_L=25$ fs.

Further, the effect of upward density ramp on self-focusing and guiding of ultra-short laser pulse is also studied. In various reports it has been shown that upward density ramps or any axial density in-homogeneity helps in stable propagation and enhancement in the wakefield acceleration [48, 49]. With upward plasma density i.e. with higher plasma density, the function $I(\xi)$ in Eq. (13) becomes steeper towards the trailing end of the pulse and also the value of $\xi c$ increases i.e. larger part of the pulse get focused and guided. Fig. 5 shows the effect of various ramp density profiles on propagation of ultra-short laser pulses for optimum laser and plasma conditions: $a_0=0.9$, $r_0=20$ μm, $\tau_L=25$ fs. To start with we have considered a density ramp profile of $n_e=n_0 \tan (z/d)$, where d is a dimensionless number which decides the steepness of the ramp profile and $n_0$ is the initial density. With $n_0=3.5\times10^{18}$ cm$^{-3}$ ($P/P_c=2.56$) and d=10, diffraction of the laser pulse is observed. This is because at z=0, the pulse sees zero density and does not have the necessary $P_c$ to get focused even within $3Z_R$. Next, we have used a slightly modified density ramp profile of $n_e=n_0+n_0 \tan (z/d)$, with $n_0=3.5\times10^{18}$ cm$^{-3}$ and d=10. We observed the pulse is more tightly focused upto 0.4-0.5 times the initial spot radius and stable propagation is also observed. Next we use a slightly modified density profile, $n_e= n_0+n_1 \tan (z/d)$ where $n_0 =3.5\times10^{18}$ cm$^{-3}$, $n_1=2n_0$ and d=10, which makes the profile comparatively steeper than the second one. With this profile although the pulse is more tightly focused a large number of oscillations occurs and makes the propagation unstable. Hence, it is observed that, presence of suitable upward density ramp enhances the self-focusing of the ultra-short laser pulses and stable propagation upto several Rayleigh lengths occur.

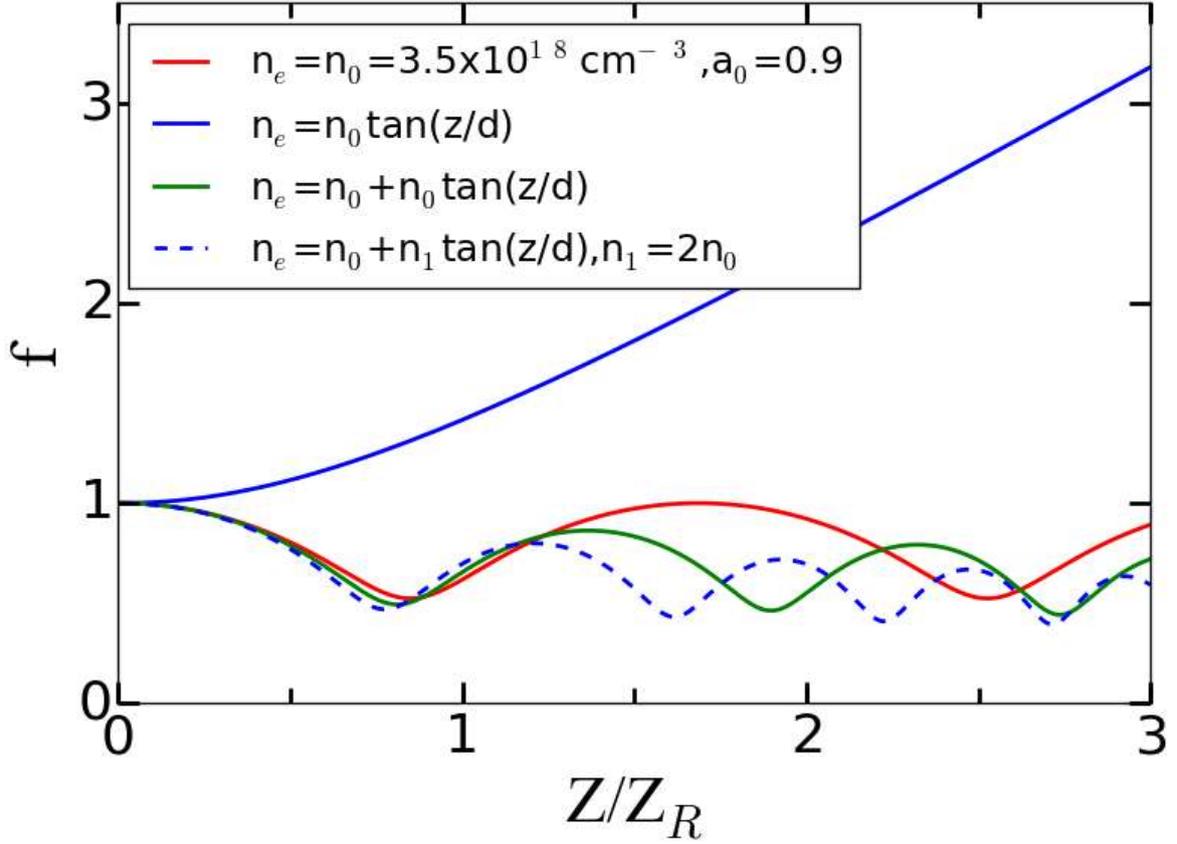

Fig.5. Evolution of f with normalized propagation distance (Z/$Z_R$) with various ramp density profiles (d=10) for fixed $n_e$=3.5×$10^{18}$cm$^{-3}$, $a_0$=0.9, $r_0$=20 μm and $\tau_L$=25 fs.

Next, a much more detailed parametric optimization study on effect of density profile steepness has been performed, considering $n_e$=$n_0$+$n_0$ tan (z/d) as the suitable ramp profile for propagation of the ultra-short laser pulses. Various density ramp profiles varying value of d from d=3 to d=12 is shown in Fig. 6a (steeper density profile for lower values of d) for initial density of $n_0$=3.5×$10^{18}$ cm$^{-3}$. Corresponding laser propagation for these density profiles are shown in Fig. 6b. It is observed that as the density profile becomes steeper, the pulse gets more tightly focused and the number of oscillation increases. A stable propagation was observed for ramp profile with d=10 for which the density increases to ~1.2 times the initial density within 3$Z_R$.

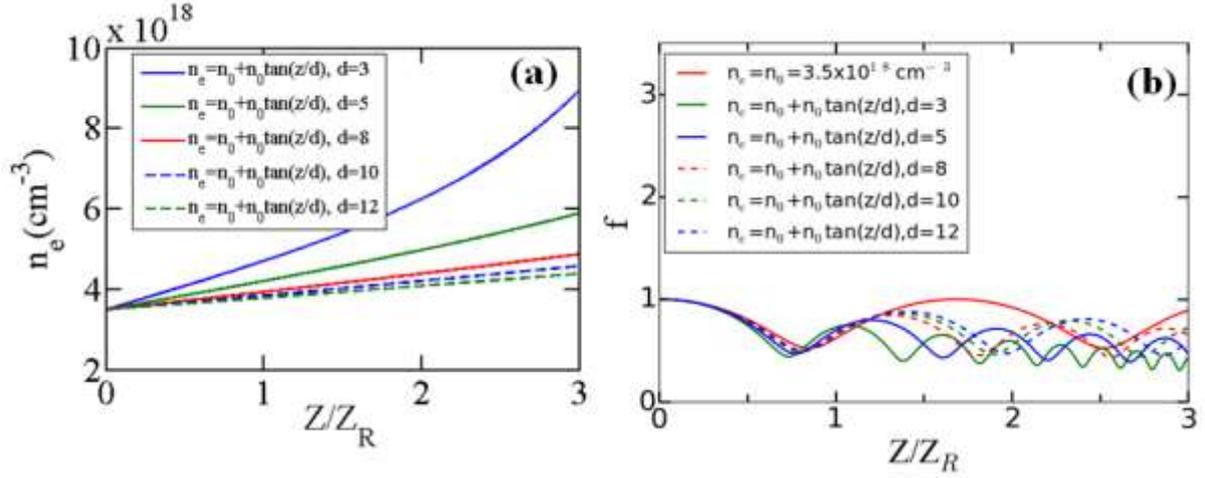

Fig.6. (a) Variation of density ramp for different d=3, 5, 8, 10, 12. (b) Optimization of ramp density profiles suitable for $L/\lambda_p \sim 0.42$ laser pulse, $a_0=0.9$, $r_0=20$ μm with d=3, 5, 8,

Further, as presence of upward density ramp enhances the self-focusing effect, its impact on the required minimum laser intensity was also studied. In Fig. 7, the pulse evolution with optimized ramp density profile $n_e=n_0+n_0 \tan(z/d)$ with $n_0=3.5\times10^{18}$ cm$^{-3}$, d=10, $r_0=20$ μm and $\tau_L=25$ fs is shown where the effect of variation of $a_0$ on propagation is studied by varying $a_0=0.4$ to 0.8. This may be compared with the effect of value of $a_0$ in case of the uniform density profile (Fig. 3). As seen effect of self-focusing and stable guiding is observed at $a_0 \geq 0.5$ with upward density ramp, compared to the value of $a_0 \geq 0.9$ for uniform density profile. This is in corroboration with the theoretical prediction. Since the initial threshold intensity for self-focusing reduces, the laser could stably guide through the plasma for a longer distance maintaining its laser intensity compared with the uniform density plasma even if loss through pump depletion be equal in both the cases. Thereby it helps in compensating the pump depletion loss associated with propagation of the pulse. Hence from this study we conclude that, besides enhancing self-focusing and stable propagation, use of a suitable ramp profile also reduces required minimum laser intensity.

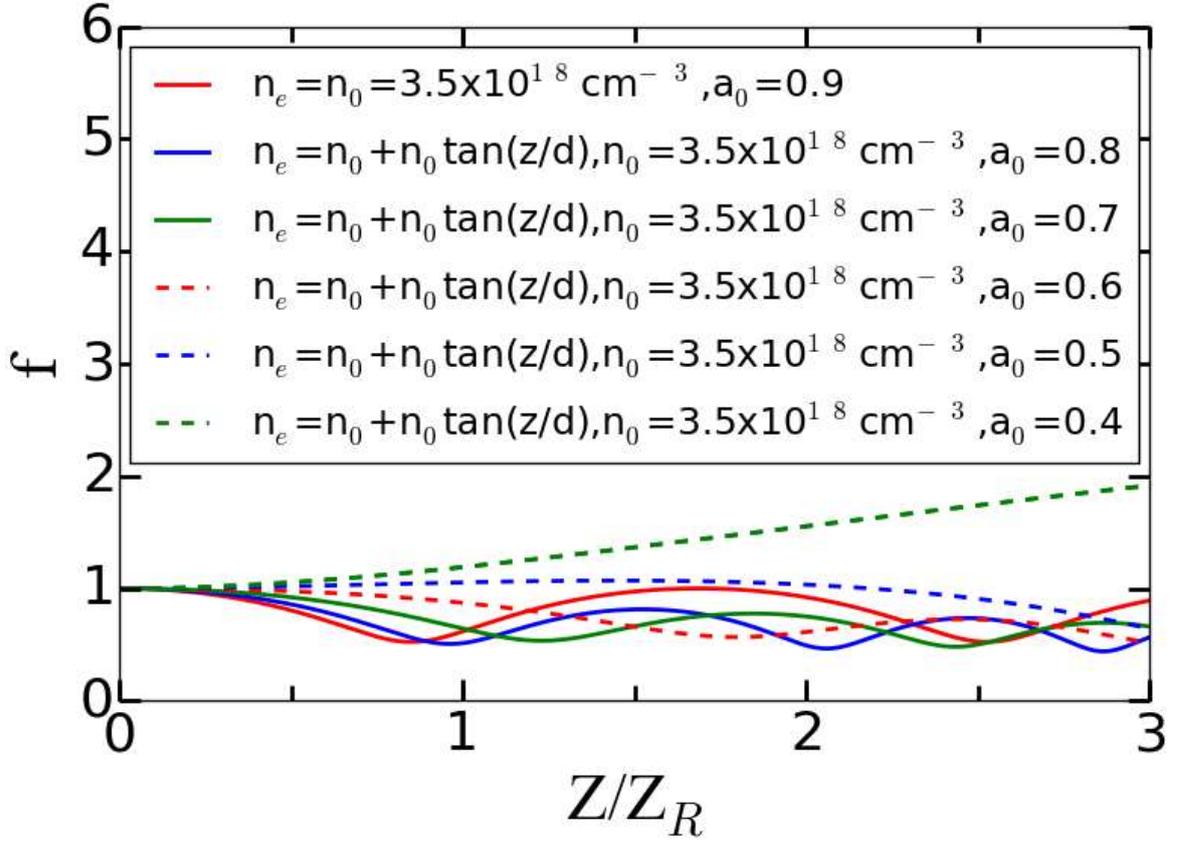

Fig.7. Evolution of f with normalized propagation distance (Z/$Z_R$) with ramp density profile $n_e = n_0 + n_0 \tan(z/d)$, d=10 for different intensities and comparison with constant density profiles.

## 4. CONCLUSION

In conclusion, numerical study of the self- focusing and guiding of ultra-short (L<$\lambda_p$) laser pulses in underdense plasma was performed by numerically solving laser spot size variation with the propagation distance where effect of plasma wakefield inhomogeneity are also considered along with relativistic effects. It was found that an ultra-short laser pulse of $L/\lambda_p$ ~0.42, an initial laser intensity of $a_0$ ~0.9, initial laser spot radius $r_0$=20 μm and $P/P_c$ ~2.56 shows the most stable propagation. Theoretical study also shows that for ultra-short laser pulse propagation, additional requirement of $r_0 \geq \lambda_p$ must also be satisfied. Further, effect of various upward density ramps on propagation of ultra-short laser pulses has also been studied. Enhancement in self-focusing together with stable propagation of the ultra-short laser pulse was observed for suitable upward density ramp profile of $n_0+n_0 \tan(z/d)$, d=10 where the density increases to 1.2 times within $3Z_R$. Most significantly, it was also shown that upward density ramp helps in reducing the required minimum laser intensity from $a_0$=0.9 to ~$a_0$=0.5 for stable guiding of the ultra-short laser pulses, thereby helps in compensating for the pump depletion loss of the pulse. Such parametric optimization studies would be helpful in generating efficient wakefields and to design laser wakefield experiments.